\documentclass[12pt]{spieman}  % 12pt font required by SPIE;
\newcommand{\ts}[1]{\textsubscript{#1}}

\usepackage{amsmath,amsfonts,amssymb}
\usepackage{graphicx}
\usepackage{setspace}
\usepackage{tocloft}
\usepackage{siunitx}
\usepackage{braket}

\title{Entanglement-based quantum key distribution with a blinking-free quantum dot operated at a temperature up to \SI{20}{K} }

\author[a,*]{Christian Schimpf}
\author[a,*]{Santanu Manna}
\author[a]{Saimon F. Covre da Silva}
\author[a]{Maximilian Aigner}
\author[a]{Armando Rastelli}
\affil[a]{Institute of Semiconductor and Solid State Physics, Johannes Kepler University, Altenbergerstraße 69, Linz, Austria}

\cftpagenumbersoff{figure}
\cftpagenumbersoff{table} 
\begin{document}

\maketitle

\begin{abstract}
Entanglement-based quantum key distribution promises enhanced robustness against eavesdropping and compatibility with future quantum networks. Among other sources, semiconductor quantum dots can generate polarization-entangled photon pairs with near-unity entanglement fidelity and a multi-photon emission probability close to zero even at maximum brightness. These properties have been demonstrated under resonant two-photon excitation (TPE) and at operation temperatures between 4 and \SI{8}{K}. However, source blinking is often reported under TPE conditions, limiting the maximum achievable photon rate. In addition, operation temperatures reachable with compact cryo-coolers could facilitate the widespread deployment of quantum dots, e.g. in satellite-based quantum communication. Here we demonstrate blinking-free emission of highly entangled photon pairs from GaAs quantum dots embedded in a p-i-n diode. High fidelity entanglement persists at temperatures of at least \SI{20}{K}, which we use to implement fiber-based quantum key distribution between two building with an average key rate of \SI{55}{bits/s} and a qubit error rate of \SI{8.4}{\percent}. We are confident that by combining electrical control with already demonstrated photonic and strain engineering, quantum dots will keep approaching the ideal source of entangled photons for real world applications.
\end{abstract}

% Include a list of up to six keywords after the abstract

% Include email contact information for corresponding author
{\noindent \footnotesize\textbf{*}Christian Schimpf,  \linkable{christian.schimpf@jku.at} \\ }
{\noindent \footnotesize\textbf{*}Santanu Manna,  \linkable{santanu.manna@jku.at} \\ }
{\noindent \footnotesize\textbf{*}These authors contributed equally.}

%------------JUST FOR DRAFTING-------------
\begin{spacing}{2}   % use double spacing for rest of manuscript
%\twocolumn
%------------------------------------------

%---------------------------------------------------------------
%           I N T R O D U C T I O N
%---------------------------------------------------------------

\section{Introduction}
\label{sect:intro}  % \label{} allows reference to this section

Quantum key distribution (QKD) relying on single photons is regarded as one of the most mature quantum technologies \cite{Pirandola2020,Ma2020}. However, the impossibility of amplifying single photons sets restrictions to the transmission distance. Entanglement-based QKD schemes are able to overcome these range-limitations when embedded in quantum networks\cite{Gisin2007,Kimble2008}, while also exhibiting a lower vulnerability to eavesdropping attacks\cite{Ekert1991,Bennett1992,Acin2007,Gisin2002,Pirandola2020}. For both fiber-based \cite{Wengerowsky2020} and satellite-based \cite{Yin2020} quantum cryptography, the most prominent sources of entangled photon pairs to date are based on the spontaneous parametric down conversion (SPDC) process. These sources are commercially available and can be operated in a large temperature range\cite{Pan2021SPDC}. As a drawback, SPDC sources exhibit approximately Poissonian photon pair emission statistics \cite{Schneeloch_2019}, which severely limits their brightness when a high degree of entanglement - and thus a low qubit error rate (QBER) - is demanded. The biexciton-exciton (XX-X) spontaneous decay cascade in epitaxially grown semiconductor quantum dots (QDs) has been demonstrated to be a viable alternative to SPDC sources due to the sub-Poissonian entangled photon pair emission statistics\cite{Benson2000}. In particular, GaAs QDs obtained by the Al droplet etching technique\cite{Gurioli2019} are capable of emitting polarization-entangled photon pairs with a fidelity to the $\ket{\phi^+}$ Bell state beyond 0.98 \cite{Huber2018,Schimpf2021}, owing to an intrinsically low exciton fine structure splitting (FSS)\cite{Huo2013} and lifetime and a near-zero multi-photon emission probability even at maximum brightness\cite{Schweickert2018}. This allowed the demonstration of entanglement-based QKD with a QBER as low as \SI{1.9}{\percent} \cite{BassoBasset2021,Schimpf2021}.

Independent of the used QD materials, the best performance in terms of entanglement fidelity and biexciton state-preparation efficiency have been obtained by operating the QD sources at temperatures below \SI{10}{K} and by using the resonant two-photon-excitation (TPE) condition \cite{Hafenbrak2007,Muller2014}. One of the drawbacks of the very low working temperatures is that they are difficult to achieve in satellites, where strong payload restrictions have to be met. Starting from about \SI{30}{K}, cryo-coolers for lower temperatures become exceedingly bulkier than higher temperature models\cite{CryoEng}. Regarding TPE, and in general resonant excitation schemes, their main limitation is represented by random charge capture in the QD, normally resulting in a significant drop of the time fraction $\beta$ in which the QD is optically active (also known as blinking) \cite{Jahn2015,Jons2017}. For the generation of single photons via strictly-resonant excitation, blinking has been successfully suppressed by embedding QDs into charge-tunable devices \cite{Warburton2000}, which allow the charge state of a QD to be deterministically controlled \cite{Zhai2020,Somaschi2016}. 
The questions whether blinking can be suppressed under TPE is not trivial, since - different from single photon resonant excitation - TPE requires excitation powers about 50 times larger, which can produce free carriers in the continuum via two-photon absorption in the barrier material.
Although two-photon absorption by a QD in a diode structure has been observed via photo-current measurements \cite{Stufler2006}, the usefulness of charge-tunable devices in the context of entangled-photon-pair generation with the TPE method has not been tested. 

Here we investigate the optical properties of GaAs QDs embedded in a p-i-n tunnel diode at a temperature of at least \SI{20}{K}, demonstrate blinking-free emission of entangled photon pairs under TPE, and use these photons to successfully implement the BBM92 QKD protocol\cite{Bennett1992} under these conditions. The key generation happens between two buildings, connected by a \SI{350}{m} long single-mode fiber inside the campus of the Johannes Kepler University.

Varying the voltage applied to the diode within a certain window allows for a fine-tuning of the emission wavelength in a range of about \SI{0.2}{nm}, while keeping the blinking-free emission intact. These structures can therefore be vital for applications in quantum networks\cite{Gisin2007}, where a precise matching of the wavelengths of multiple emitters becomes important\cite{Giesz2015,Reindl2017} and the transmission rate scales with $\beta^2$\cite{Jons2017}.

%---------------------------------------------------------------
%          T U N N E L   D I O D E
%---------------------------------------------------------------

\section{GaAs Quantum Dots in a Tunnel Diode Structure at a Temperature of \SI{20}{K}}

\begin{figure*}
\begin{center}
\begin{tabular}{c}
\includegraphics[]{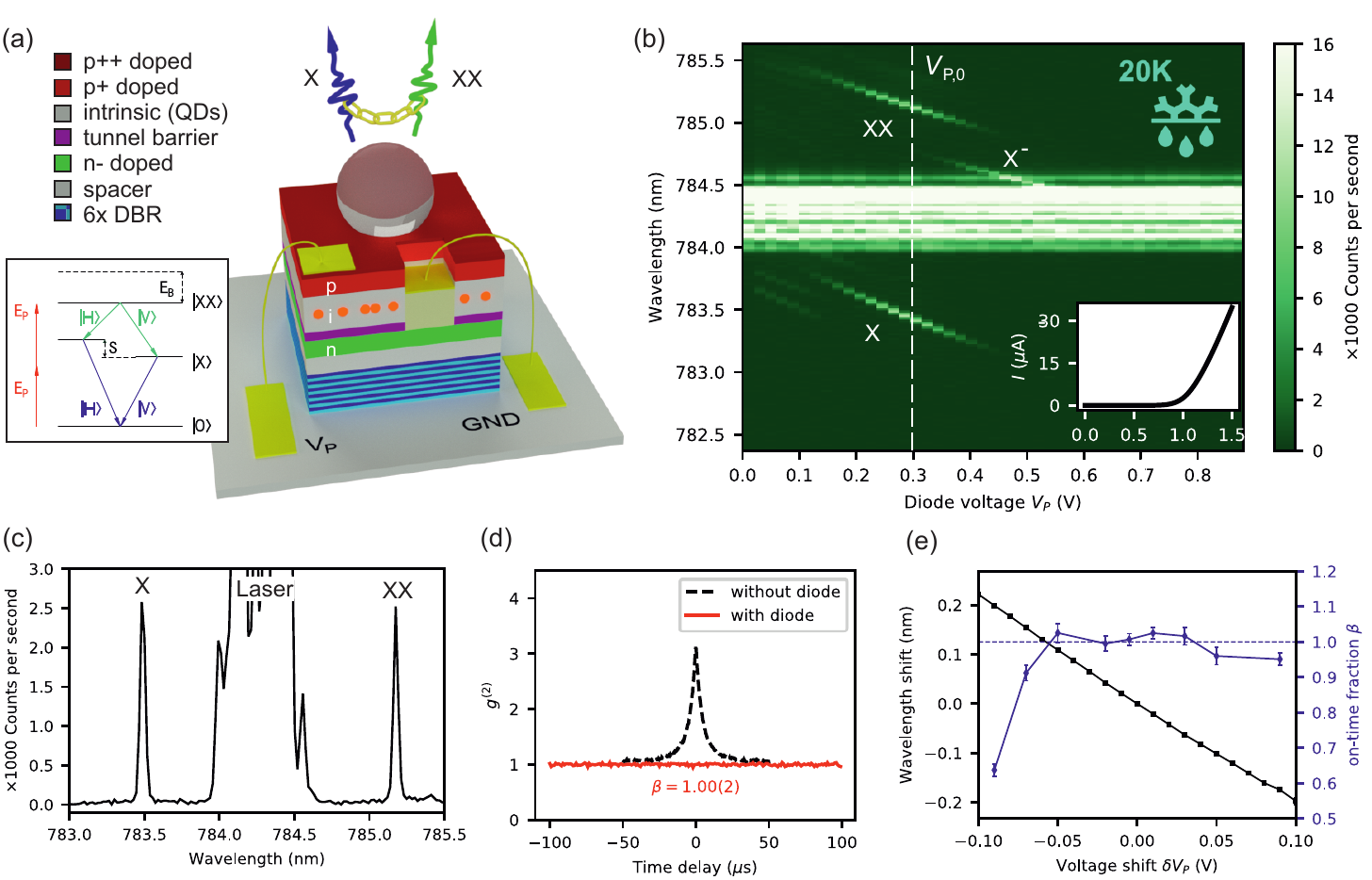}
\end{tabular}
\end{center}
\caption 
{ \label{fig:PL}
Photoluminescence properties of GaAs quantum dots (QDs) in a p-i-n diode structure at a temperature of \SI{20}{K}, excited by resonant two-photon excitation (TPE). (a) p-i-n diode structure with a tunnel barrier between the n-doped and the intrinsic region. The inset shows the principle of TPE, with $E_\text{P}$ the laser energy, $E_\text{B}$ the biexciton (XX) binding energy and $S$ the exciton (X) fine structure splitting. (b) Emission spectra at TPE conditions when sweeping the diode voltage $V_\text{P}$ in forward bias. The inset shows the diode current $I$ over $V_\text{P}$. The white dashed line indicates $V_{\text{P,0}}=\SI{0.3}{V}$, at which the diode is operated during the QKD experiment. (c) Emission at $V_\text{P}=V_{\text{P,0}}$. (d) Second order correlation function $g^{(2)}$ of the X signal with a time-bin of \SI{1}{\micro s} at $V_\text{P}=V_{\text{P,0}}$. The  $g^{(2)}$ is shown for the QD in the diode structure (red), indicating an on-time fraction of $\beta=1.00(2)$ and a QD without diode (black, dashed) with a typical value of $\beta\approx 0.3$. (e) Wavelength shift and $\beta$ for different deviations $\delta V_\text{P}=V_{\text{P}}-V_{\text{P,0}}$. The blue dashed lines indicates a value of $\beta$ corresponding to no blinking.} 
\end{figure*} 

The GaAs QDs in a p-i-n tunnel diode structure, as depicted in Fig. \ref{fig:PL}(a), were grown by Molecular Beam Epitaxy (MBE). The first functional layers are six pairs of Distributed Bragg Reflectors (DBRs), grown by alternating Al\ts{0.95}Ga\ts{0.05}As and Al\ts{0.20}Ga\ts{0.80}As layers with \SI{65.2}{nm} and \SI{56.6}{nm} thickness, respectively. The n-doped region is formed by a \SI{95}{nm} thick Al\ts{0.15}Ga\ts{0.85}As layer with an Si concentration of \SI{E18}{cm^{-3}}. A combination of layers (\SI{15}{nm} Al\ts{0.15}Ga\ts{0.85}As plus \SI{8}{nm} Al\ts{0.33}Ga\ts{0.67}As) acts as tunnel barrier between the n-layer and the QDs, which are obtained via the local Al droplet etching technique\cite{Gurioli2019} and covered by the \SI{268}{nm} thick Al\ts{0.33}Ga\ts{0.67}As intrinsic region. A \SI{65}{nm} layer of Al\ts{0.15}Ga\ts{0.85}As, doped with \SI{5E18}{cm^{-3}} carbon atoms forms the p-region. Layers of \SI{5}{nm} Al\ts{0.15}Ga\ts{0.85}As and \SI{10}{nm} GaAs, each doped with \SI{E19}{cm^{-3}} carbon, cap the structure in order to form a conductive surface and to protect the active region from oxidation. The DC voltage $V_\text{P}$ is applied to the p-contact with respect to the grounded n-contact (see supplementary for details about the electrical contacts). A solid immersion lens on top of the structure enhances the overall extraction efficiency to about \SI{3}{\percent}.

The sample containing the GaAs QDs is fixed with silver-glue on the cold-finger of a He flow cryostat and cooled to a temperature of \SI{20}{K}. The temperature is measured with a calibrated silicon diode placed under the cold-finger, so we estimate that the actual sample temperature could be up to \SI{5}{K} higher. One individual QD is optically excited by a focused  pulsed laser with a repetition rate of \SI{80}{MHz} and the laser energy $E_\text{P}$ tuned to half of the biexciton (XX) energy, as depicted in the inset of Fig. \ref{fig:PL}(a). This resonant two-photon excitation (TPE) scheme is the same as used in previous experiments\cite{Schweickert2018,Huber2018,Schimpf2021}.
Figure \ref{fig:PL}(b) depicts the micro-photoluminescence spectra of a QD when adjusting $E_\text{P}$ at the working point $V_\text{P,0}=\SI{0.3}{V}$, used for the further measurements, and then sweeping $V_\text{P}$ from 0 to \SI{1.5}{V}. The inset shows the corresponding diode current. In the voltage range of about \SI{0.15}{V} to \SI{0.35} the QD is in its charge neutral configuration and only the biexciton (XX) and the exciton (X) transition lines are visible. For higher voltages, a single electron tunnels into the QD so that the negative trion (X\textsuperscript{-}) is addressed via phonon-assisted excitation. Figure \ref{fig:PL}(b) shows the spectrum at $V_\text{P,0}$, with the XX excited at pi-pulse. At this voltage, the auto-correlation function $g^{(2)}$ for a time span of \SI{100}{\micro s} and a time-bin of \SI{1}{\micro s} was recorded, depicted as the red data points in Fig. \ref{fig:PL}(d) (In this case, the time-bin is much larger than the excitation period of \SI{12.5}{ns}, so that the anti-bunching from the single-photon emission becomes invisible). The $g^{(2)}$ is one for all times delays, which indicates the complete absence of blinking, i.e., an on-time fraction of $\beta=1/g^{(2)}(0)=1.00(2)$\cite{Jahn2015}. The black dashed line indicates the $g^{(2)}$ typically measured for previously used GaAs QDs without diode, resulting in $\beta\approx0.3$.

It is interesting to note that the tunnel provides a window for $V_\text{P}$ of about \SI{0.1}{V} in which the QDs can be operated without blinking. This offers a tuning range for the central emission wavelengths of a about \SI{0.2}{nm} as depicted for a different, but representative QD in Fig. \ref{fig:PL}(e). The blue dashed line indicates the case of $\beta=1$, corresponding to no blinking. The same tuning-range was observed also for temperatures well below \SI{20}{K}.

%---------------------------------------------------------------
%          E N T A N G L E M E N T
%---------------------------------------------------------------

\section{Characterization of the Entangled Photon Pairs}

\begin{figure*}
\begin{center}
\begin{tabular}{c}
\includegraphics[]{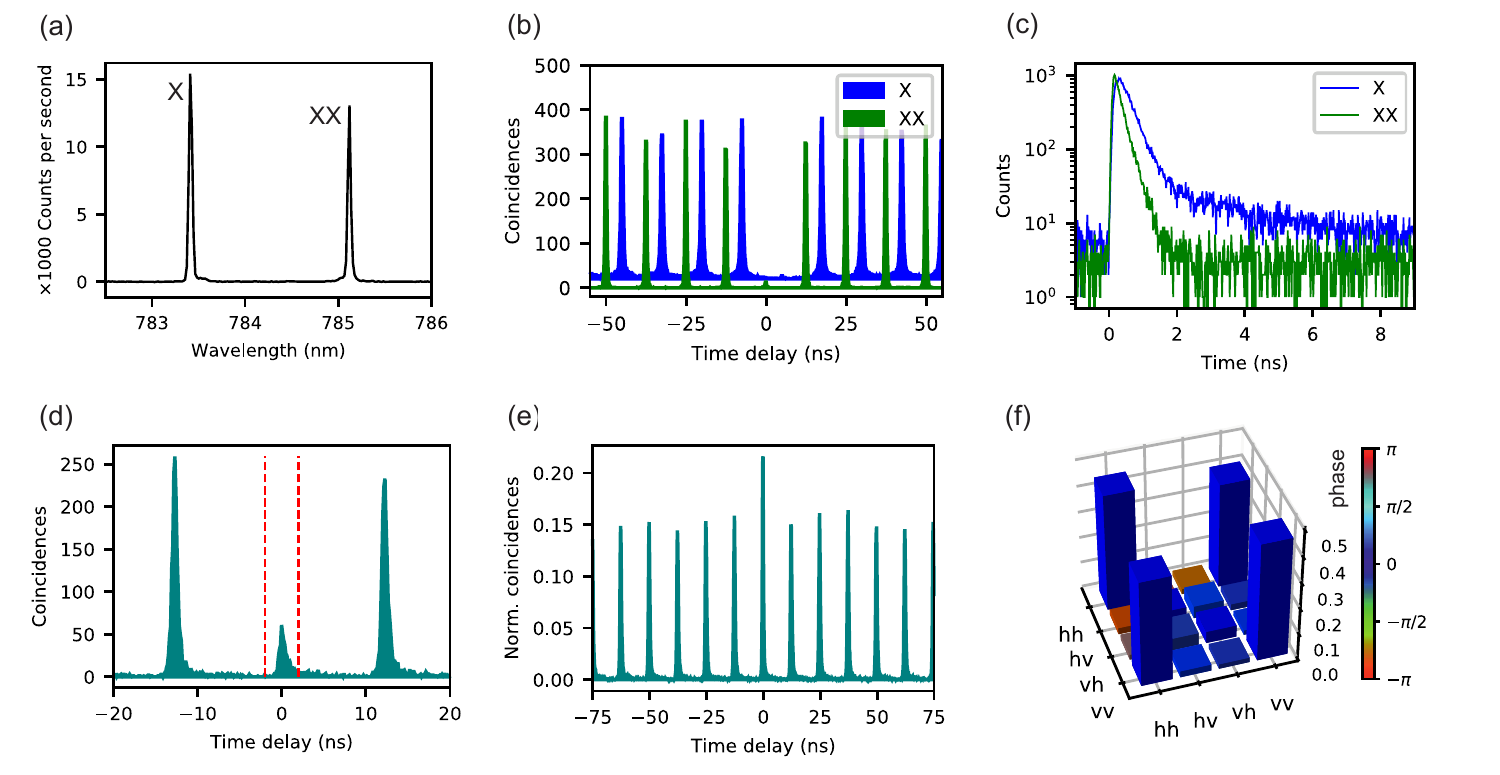}
\end{tabular}
\end{center}
\caption 
{ \label{fig:Entanglement}
Emission properties relevant for the polarization entanglement, measured at a temperature of $\SI{20}{K}$. (a) Spectra of the individually filtered XX and X emission lines, combined at a 50:50 fiber beam splitter. (b) Single-photon emission characteristics of the XX and X signals, observed by detecting coincidences in a Hanbury-Brown-Twiss arrangement. (c) Lifetime trace of the XX and X signals. The X signal exhibits a slow secondary decay channel, which is not present at temperatures lower than \SI{10}{K}. (d) Example among the 36 recorded coincidence histograms between the XX and X detection, corresponding to a measurement in the HV basis. The red dashed lines indicate the time-bin of \SI{2}{ns} in which the coincidences are summed up to calculate the peak areas. (e) Unpolarized coincidence measurement between the XX and X photons. The excess coincidences at zero time delay stem from a non-unity photon-pair generation probability. (f) Density matrix of the two-photon polarization entangled state of the XX and X photons, recorded by full state tomography.} 
\end{figure*} 

The QDs of the here used diode sample exhibit an average FSS of about \SI{6}{\micro eV}. Average values below \SI{4}{\micro eV} are regularly achieved for nano-holes created at a substrate temperature of \SI{600}{\degree C} \cite{Huo2013,Keil2017}, while a slightly higher temperature (about \SI{610}{\degree C}) was used here to possibly improve the crystal quality. Instead of resorting to strain tuning to bring the FSS of one individual QD to zero\cite{Huber2018}, the stochastic distribution of the FSS among all QDs on the sample can be used to find a QD with a suitably low FSS for the given use case. In order to estimate the FSS required for a serviceable QBER in a BBM92 QKD arrangement we first calculate the 2-qubit density matrix in polarization space\cite{Hudson2007}, considering only the FSS and the typically observed X lifetime of about \SI{230}{ps}. From the density matrix $\rho$, we calculate the expected QBER via
\begin{equation}
    q=\frac{1}{2}\sum \limits_{i=1}^{4} \braket{O_i|\rho|O_i},
\label{eq:qber}
\end{equation}
where $O_i \in \{H_\text{A} V_\text{B}, V_\text{A} H_\text{B}, D_\text{A} A_\text{B}, A_\text{A} D_\text{B}\}$ are the 2-qubit measurement bases between Alice (A) and Bob (B), in which a measured coincidence corresponds to a false key bit. After a brief scanning we chose a QD with an FSS of \SI{0.96(9)}{\micro eV}, for which we calculate a minimum QBER of \SI{2.7}{\percent}.

The XX and the X photons generated by TPE are filtered out individually and coupled into single mode fibers. Figure \ref{fig:Entanglement}(a) shows the spectra of the XX and X emission lines merged at a 50:50 fiber beam splitter, of which one output is sent to the spectrometer (The slightly lower XX signal intensity stems from the higher distance from the objective to the fiber collimator compared to the X signal, resulting in a lower coupling efficiency).

Before performing the QKD experiment, a characterization of the single-photon emission characteristics and the polarization entanglement between the XX and the X photons is performed, as those properties primarily determine the QBER during the key generation process. The most important quantities are summarized in Table \ref{tab:performance} and compared to the values for a different QD (in the same diode structure), measured at $\SI{5}{K}$. Figure \ref{fig:Entanglement}(b) depicts a coincidence histogram of an auto-correlation measurement for both the XX and the X signals. Using a time-bin of $\SI{2}{ns}$, the results are $g^{(2)}_\text{XX}(0)=0.034(4)$ and  $g^{(2)}_\text{X}(0)=0.020(3)$, which are comparable to previously measured values using the same optical arrangement at a temperature of \SI{5}{K} (see Table \ref{tab:performance}), but with a slightly higher value for the XX signal. Compared to data acquired at \SI{5}{K}, one can observe a broadening of the peaks belonging to the X signal. Inspecting the lifetime traces, which are shown for both the XX and the X in Fig. \ref{fig:PL}(c), it becomes evident that this broadening stems from a slow decay channel of the X, which overlays with the mono-exponential decay from the bright X to ground state usually observed at \SI{5}{K} (see supplementary for lifetime- and cross-correlation measurements at \SI{5}{K}). A convoluted fit results in an X lifetime of $T_{1,\text{X}}=\SI{252(9)}{ps}$, with the caveat that this value is slightly overestimated due to the presence of the slow decay channel. The XX lifetime is measured as $T_{1,\text{XX}}=\SI{72(3)}{ps}$, which is significantly lower than the \SI{120}{ps} typically observed at $\SI{5}{K}$. These temperature-dependent phenomena are visible in the lifetime traces are subject of further investigation, as they could shed light on the changing dynamics of quasi particles in GaAs QDs at higher temperatures.

A full state tomography\cite{James2001} is performed to determine the degree of entanglement between the XX and the X photons. Figure \ref{fig:Entanglement}(d) shows an example among the 36 recorded XX/X coincidence histograms, corresponding to a measurement in the HV basis. The red dashed lines indicate the time-bin of \SI{2}{ns} in which the coincidences are summed up and compared with the average side peak area. The coincidences stemming from the slow X decay channel (the "tail" on the right side of the peaks), which are partially excluded by this time-bin, account for about \SI{9}{\percent} of the total coincidences per excitation cycle. Figure \ref{fig:Entanglement}(e) depicts the weighted sum of the histograms corresponding to the measurement bases HH, VV, HV and VH (with H the horizontal and V the vertical polarization), normalized by their respective average side peak areas. From this histogram a photon pair generation probability per excitation pulse of $\epsilon=0.87(2)$ is calculated (see supplementary for details), which is marginally lower than the value of $0.91(2)$ observed at \SI{5}{K}.

From the 36 correlation histograms the 2-qubit density matrix in polarization space is calculated and depicted in Fig. \ref{fig:Entanglement}(f). The maximum likelihood estimator used during this process is adapted for the dynamics of the QD light emission (see supplementary). The derived concurrence is $0.713(8)$ and the maximum fidelity to a Bell state is $0.925(5)$, which show a significant drop compared to the values obtained at \SI{5}{K}. However, the expected QBER of \SI{7.45}{\percent} calculated by Eq. \eqref{eq:qber} is still well below the $\SI{11}{\percent}$ allowed for the BBM92 protocol\cite{Bennett1992,Gisin2002}, and therefore suitable for performing QKD.

\begin{table*}[ht]
\caption{Summarized emitter performance for two representative QDs in a tunnel diode structure excited by TPE, measured at temperatures of \SI{5}{K} and \SI{20}{K}, respectively.}
\label{tab:performance}
\begin{center}       
\begin{tabular}{|l||c|c||c|c|} %% this creates two columns
%% |l|l| to left justify each column entry
%% |c|c| to center each column entry
%% use of \rule[]{}{} below opens up each row
\hline
\rule[-1ex]{0pt}{3.5ex}  Temperature &  \multicolumn{2}{c||}{\SI{5}{K}} & \multicolumn{2}{c|}{\SI{20}{K}} \\
\hline
\rule[-1ex]{0pt}{3.5ex}  & X & XX & X & XX \\
\hline\hline
\rule[-1ex]{0pt}{3.5ex}  $g^{(2)}(0)$ & 0.017(4) & 0.011(3) & 0.020(3) & 0.034(4) \\
\hline
\rule[-1ex]{0pt}{3.5ex}  Lifetime (ps)  & 238(3) & 116(2) & 252(9) & 72(3)\\
\hline
\rule[-1ex]{0pt}{3.5ex}  Pair generation efficiency & \multicolumn{2}{c||}{0.91(2)} &  \multicolumn{2}{c|}{0.87(2)} \\
\hline
\rule[-1ex]{0pt}{3.5ex}  FSS (\SI{}{\micro eV}) & \multicolumn{2}{c||}{1.13(7)} &  \multicolumn{2}{c|}{0.96(9)}\\
\hline
\rule[-1ex]{0pt}{3.5ex}  Calculated Concurrence\textsuperscript{*} & \multicolumn{2}{c||}{0.905} &  \multicolumn{2}{c|}{0.900}\\
\hline
\rule[-1ex]{0pt}{3.5ex}  Measured Concurrence & \multicolumn{2}{c||}{0.904(3)} &  \multicolumn{2}{c|}{0.713(8)}\\
\hline
\rule[-1ex]{0pt}{3.5ex}  Calculated Fidelity to $\ket{\phi^+}$\textsuperscript{*} & \multicolumn{2}{c||}{0.959}&  \multicolumn{2}{c|}{0.960} \\
\hline
\rule[-1ex]{0pt}{3.5ex}  Measured Fidelity to $\ket{\phi^+}$ & \multicolumn{2}{c||}{0.975(1)}&  \multicolumn{2}{c|}{0.925(2)} \\
\hline
\multicolumn{5}{l}{\footnotesize{\textsuperscript{*}Only considering expectation values of measured $g^{(2)}$, X lifetime and FSS\cite{Hudson2007}}} 
\end{tabular}
\end{center}
\end{table*}

%---------------------------------------------------------------
%          Q K D
%---------------------------------------------------------------

\section{Quantum Key Distribution with Entangled Photons}

\begin{figure}
\begin{center}
\begin{tabular}{c}
\includegraphics[]{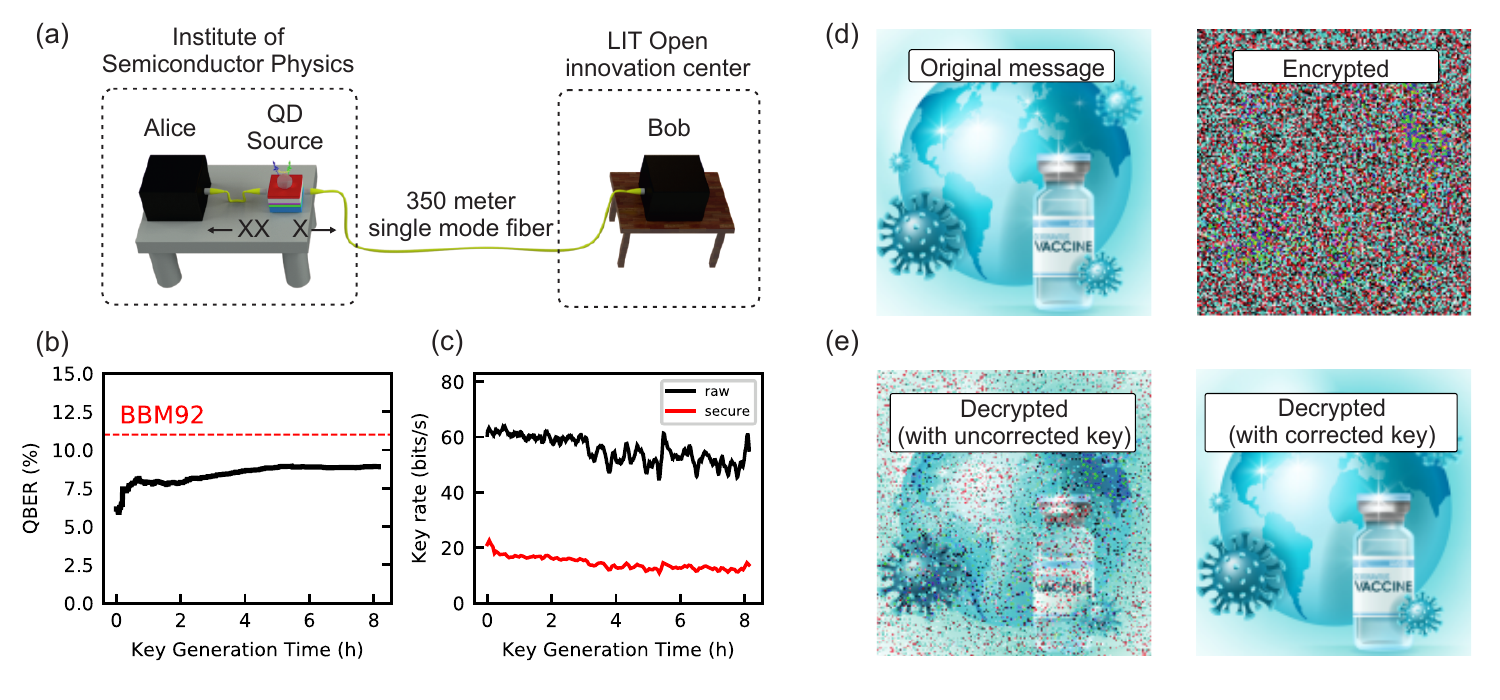}
\end{tabular}
\end{center}
\caption 
{ \label{fig:KeyGen}
Key generation in the BBM92 protocol over a time span of about 8 hours. (a) QKD arangement. Alice and the photon source are situated on an optical table, Bob is placed in a movable box on a table in another building and connected with the source via a \SI{350}{m} long single mode fiber. (b) Qubit error rate (QBER) during the key generation with an average of \SI{8.42}{\percent}. (c) Raw (black) and secure (red) key rate during the key generation with average values of \SI{54.8}{bits/s} and \SI{14.2}{bits/s}, respectively. (d) Encryption of a bitmap with the dimensions of $144\times 150$ pixels and a color-depth of 24 bit, resulting in a total size of \SI{64.8}{kbyte}. The encryption with Alice's key yields a scrambled message ready to be sent over a public channel. (e) Decryption at Bob's site when using an uncorrected key (left) and a corrected key (right).} 
\end{figure} 

The XX and X photons in their individual single mode fibers are distributed to Alice and Bob, as sketched in Fig. \ref{fig:KeyGen}(a). The infrastructure is identical to the one used in a previous QKD experiment\cite{Schimpf2021}. After an initial synchronization and polarization correction routine, the key generation is performed overnight for a total duration of about 8 hours. The observed QBER, depicted in Fig. \ref{fig:KeyGen}(b), has an average value of \SI{8.42}{\percent}. 

The QBER shows a minimum of \SI{6.15}{\percent}, which is even lower than the \SI{7.45}{\percent} estimated from $\rho$, shown in Fig. \ref{fig:Entanglement}(f). This discrepancy probably arises from the way the time-synchronization between Alice and Bob is maintained: In order to dynamically adjust the relative time delay between the arrival times of the XX and X photon at Alice and Bob, respectively, a peak in the continuously measured cross-correlation function is tracked. All photons around the peak-maximum in a time window of \SI{2}{ns} are used for key generation. We found that already a shift of the center of this time-bin by \SI{0.5}{ns} (about the time-resolution of our single photon detectors) can change the QBER by around \SI{2}{\percent}, because a higher contribution of photons stemming from the slow X decay channel, shown in Fig. \ref{fig:Entanglement}(c), decreases the entanglement (see supplementary for details). The time-window during the key generation was therefore probably shifted with respect to the one used for calculating $\rho$. These findings are not only interesting for a better understanding of the underlying mechanisms degrading the entanglement at higher temperatures, but also indicate that with careful choice of the time-window (within the limits of the detector resolution) a compromise between QBER and efficiency during the key generation process can be made.

We attribute the changing QBER over the course of the measurement to the following factors: The FSS changed from the initial \SI{0.96(9)}{\micro eV} to \SI{1.35(11)}{\micro eV} over the course of the QKD measurements, which can add about \SI{2}{\percent} to the QBER due to an accelerated phase rotation process leading to a decreased fidelity to the ideal Bell state on average\cite{Hudson2007}. A changing FSS was never noticed before in these kinds of samples. A possible origin could be a changing strain state in the silver glue used to stick the sample on the chip carrier. The second contribution to a rising QBER could be a varying ambient temperature, that affects the polarization rotation exerted by the fibers, which then leads to a larger deviation from the $\ket{\phi^+}$ Bell state, for which our QKD setup is designed. After the initial polarization correction no active correction was performed during the 8 hours of key generation.

Figure \ref{fig:KeyGen}(c) presents the raw (black) and secure (red) key generation rate. The secure key rate depends on the raw key rate and the QBER during the respective time window, as the QBER largely determines the efficiency of the key reconciliation process as described in Ref.\cite{Schimpf2021}. A change in raw key rate over time occurs due to a slight drift of the cryostat relative to the objective, leading to a decreasing excitation- and collection-efficiency. We counter this effect by an automatized movement of the X/Y position of the cryostat via linear stages to optimize the average detector count rates, should they fall below a certain threshold. The average raw key rate over the full time span is found to be \SI{54.8}{bits/s}, with a resulting average secure key rate of \SI{14.2}{bits/s}. 

The generated (raw) key is used to encrypt a bitmap with a size of \SI{64.8}{kbyte}, see Fig. \ref{fig:KeyGen}(d), using a one-time-pad procedure. The decryption is depicted in Fig. \ref{fig:KeyGen}(e), for the cases when the raw key pair was used (left) and the corrected one (right).

%---------------------------------------------------------------
%          D I S C U S S I O N
%---------------------------------------------------------------

\section{Discussion and Conclusion}

In this work, we have demonstrated a blinking-free source of polarization-entangled photon pairs, based on a GaAs quantum dot operated at a temperature of at least \SI{20}{K}. The intrinsically low fine structure splitting, owing to the local Al droplet etching technique\cite{Gurioli2019}, together with the employed p-i-n diode allows us to generate an uninterrupted stream of photon pairs with a fidelity to the $\ket{\phi^+}$ Bell state of \SI{0.925(2)} when using the pulsed two-photon-excitation scheme\cite{Hafenbrak2007,Muller2014}. The device also allows the fine-tuning of the emission wavelength within a range of \SI{0.2}{nm} while keeping the blinking-free operation intact, which is favourable for interconnecting multiple sources to quantum networks\cite{Reindl2017,Jons2017,Kimble2008,Gisin2007}.

The source was used to demonstrate quantum key distribution (QKD) via the BBM92 protocol\cite{Bennett1992} between two parties in two different buildings of the Johannes Kepler University, connected via a \SI{350}{m} long underground single mode fiber. The average qubit error rate (QBER) was \SI{8.42}{\percent}, which is well below the threshold of \SI{11}{\percent} required for generating secure keys when using the BBM92 protocol\cite{Bennett1992,Gisin2002}. From an initial key rate of \SI{54.8}{bits/s} error-free and privacy-amplified\cite{Schimpf2021} keys could be distilled with a rate of \SI{14.2}{bits/s}.

Comparing the decay dynamics of the biexciton and exciton states at temperatures of \SI{20}{K} and \SI{5}{K} allows us to identify a secondary slow decay channel of the exciton as the major entanglement degrading mechanism. While the physical origin of this observation and the elaboration of possible solutions will require further investigations, we find that the QBER during QKD can be optimized by a mild time filtering (see supplementary). This work makes us  optimistic that by combining electrical control with advanced photonic processing \cite{Liu2019,Wang2019} and strain-tuning platforms \cite{Trotta2015,Huber2018} will lead to nearly ideal sources of entangled photon pairs that can be operated in demanding environments.

%---------------------------------------------------------------
%          A P P E N D I X
%---------------------------------------------------------------

\appendix    % this command starts appendixes
\section*{Disclosures}
The authors declare no conflict of interest

\section*{Acknowledgements}
Christian Schimpf is a recipient of a DOC Fellowship of the Austrian Academy of Sciences at the Institute of Semiconductor Physics at Johannes Kepler University, Linz, Austria. We thank C. Diskus for providing the Rb clocks used for QKD, and M. Reindl and L. Costadedoi for their assistance in the optics laboratory.

This work was financially supported by the Austrian Science Fund (FWF) via the Research Group FG5, P 29603, P 30459, I 4320, I 4380, I 3762, the European Union's Horizon 2020 research and innovation program under grant agreements No. 899814 (Qurope) and No. 871130 (Ascent+), the Linz Institute of Technology (LIT) and the LIT Secure and Correct Systems Lab, supported by the State of Upper Austria.

%%%%% References %%%%%

\bibliography{report}   % bibliography data in report.bib
\bibliographystyle{spiejour}   % makes bibtex use spiejour.bst

%%%%% Biographies of authors %%%%%

\end{spacing}
\end{document}